\documentstyle[epsfig]{mn}

\begin{document}
\title[Non-axisymmetry in advanced mergers of galaxies]
{Measurement of non-axisymmetry 
  in centres of advanced mergers of galaxies}

\author[C.J. Jog and A. Maybhate]
       {Chanda J. Jog$^{1}$\thanks{E-mail : cjjog@physics.iisc.ernet.in} and
        Aparna Maybhate$^{2}$\thanks{previously Chitre, E-mail : maybhate@stsci.
	edu; Present address: Space
Telescope Science Institute, 3700 San Martin Drive, Baltimore, MD 21218, USA}\\
$^1$   Department of Physics,
Indian Institute of Science, Bangalore 560012, India \\
$^2$Department of Astronomy and Astrophysics, 525 Davey
Laboratory, Pennsylvania State University, University Park, PA
16802, USA\\
} 


\maketitle

\begin{abstract}
   We measure the non-axisymmetry in the luminosity distribution in the inner 
   few kpc of the remnants of advanced mergers of galaxies with
a view to understand the relaxation in the central regions. For this, we 
analyze the images from the Two Micron All Sky Survey (2MASS) 
archival data for a selected sample of 12 merging galaxies, which show signs
of interaction but have a single nucleus. The central regions are fitted 
by elliptical isophotes whose centres are allowed  to vary to get the best fit.
The centres of isophotes show a striking sloshing pattern 
with a spatial variation of  up to 20-30 \% within the central 1 kpc. 
This indicates mass asymmetry and a dynamically unrelaxed behaviour 
in the central region. Next, we 
Fourier-analyze the galaxy images while keeping the centre 
constant and  measure the deviation from axisymmetry in terms of the 
 fractional Fourier amplitudes (A$_1$, A$_2$ etc) as a function of 
 radius.  All the mergers show a high value of lopsidedness 
  (upto A$_1 \sim 0.2$) in the  central  5 kpc.
 The m=2 asymmetry is even stronger, with values of A$_2$ 
upto $\sim 0.3$, and in three cases these are shown to
represent  bars.  The corresponding values denoting
non-axisymmetry in inner
regions of a control sample of eight non-merger galaxies are
found to be several times smaller. 
 
Surprisingly, this central asymmetry 
is seen even in mergers where the outer regions have relaxed
into  a smooth elliptical-like r$^{1/4}$ profile or a
spiral-like  exponential profile. Thus the 
 central asymmetry is long-lived, estimated to be $\sim$ 1 Gyr,
and hence lasts for over 100 local dynamical timescales. These central 
asymmetries are 
 expected to play a key role in the future dynamical evolution of the central 
 region of a merger, and can help in feeding of a central AGN.
\keywords}

\end{abstract}

\section{Introduction}
Interactions and mergers of galaxies are now known to be common and these 
can significantly affect the dynamics and evolution of the merging galaxies
(Wielen 1990, Barnes \& Sanders 1999). 
The work so far has mainly concentrated on the intermediate radial range of 
a few kpc - few $\times$ 10 kpc : observationally, the mergers have been shown 
to result in 
an elliptical-like $r^{1/4}$ profile as was first shown for NGC 7252 by
Schweizer (1982), and later by others for larger samples (Wright et al. 1990, 
Stanford \& Bushouse 1991, Chitre \& Jog 2002). 
Interestingly, in half the sample of mergers showing a single nucleus but
evidence for tidal interaction as studied by Chitre \& Jog (2002), the 
remnants even showed 
an exponential profile like a disc while a few showed no-fit implying 
unrelaxed outer regions. The galaxies with disc-like profiles are peculiar 
and  have elliptical-like high velocity dispersion and could have 
thick discs, as shown for two galaxies (Jog \& Chitre 2002). 
Theoretically the evolution of mergers leading to a relaxed
elliptical-like remnant has been well-studied by N-body
simulations (e.g., Barnes 1992, Bournaud et al. 
2005 b). Recent simulations of unequal-mass mergers can reproduce the above, 
relaxed disc-like remnants (Bournaud et al. 2004, Naab \&
Trujillo 2005), while the transition 
between the elliptical-like and spiral-like remnants is shown to occur for a 
narrow mass-range of 3:1-4.5:1 (Bournaud et al. 2005 b). An inclusion of gas 
has also shown to result in a disc-like remnant in some cases 
(Volker \& Hernquist 2005). 

 However despite its obvious importance for the evolution of the
central regions, 
the luminosity distribution in the central regions of mergers has not been 
studied systematically so far. A few mergers where this has been studied,
such as NGC 3921 (Schweizer 1996), and Arp 163 (Chitre \& Jog 2002),
show a wandering or meandering centre for the isophotes.
But there has been no study so far to measure the non-axisymmetric
amplitudes in mergers.
 
In this paper, we systematically study the asymmetry in the 
central few kpc  of advanced mergers of galaxies.
The recent availability of near-IR data (including the excellent archival 2MASS
database) allows one to study the main underlying mass distribution directly 
in the central regions of mergers for a  large sample of galaxies. In the 
near-IR, the dust extinction is low and the luminosity distribution directly 
traces the old stellar component, and hence the main mass component.

The mergers involve a 
patently non-axisymmetric configuration, especially at large
radii (of 20-30 kpc) outside of the main merged body where spectacular 
tidal features are seen.
Even the inner optical image is patchy, especially as seen
in the blue light where the dust obscuration plays a major role. Hence,
naively it would seem that they 
would show asymmetry in the central regions of few kpc as well.
However, the advanced mergers do show some degree of
relaxation as evident from their merged centres, and their relaxed  outer profiles ($\sim
10$ kpc or so), as seen in the near-IR, as discussed above. 
Further, the dynamical timescales are
smaller in the inner regions and hence they would be expected to
be more relaxed than the outer parts. Thus it is interesting to 
measure the values of non-axisymmetry in the central regions of mergers. 
For example, the values of central lopsidedness can provide an important 
 diagnostic for the degree of relaxation in the central
regions of a merging pair of galaxies.

Our earlier study of Arp 163 (Chitre \& Jog 2002) showed a
wandering centre. This is a no-fit merger -  which cannot be
fit by an elliptical or an exponential fit, where the outer regions
are not relaxed.
In this paper, we measure the central asymmetry in the mergers
with no-fit in the outer regions as well 
as in the other mergers which show relaxed outer profiles.  Thus this study 
is expected to provide information on the relaxation in the inner regions, 
and also reveal the correlation if any of the central asymmetry or degree 
of relaxation with that in the outer regions.

We do a detailed isophotal analysis of the galaxy images 
and study the sloshing of the isophotal centres quantitatively, and then 
Fourier analyze the images and obtain the amplitudes for lopsidedness and 
also for m=2. In three cases  the latter are shown to represent a strong bar.

We show that the lopsidedness is high in all cases,
and the magnitude of the central asymmetry is higher in those  
mergers which are unrelaxed in the outer regions. 
Interestingly, even galaxies where the outer regions have relaxed
into a  smooth mass profile still display a fairly strong central 
asymmetry and hence a non-relaxed central mass distribution.

Section~2 contains a discussion about the sample selection, and
the basic properties of the sample galaxies.  Section~3 contains 
data analysis and the results for mergers,  and a 
comparison with results for a set of non-merger or normal galaxies. 
Section~4 includes a
discussion of the dynamical implications of the results, and a brief 
summary is given in Section~5.

\bigskip

\section{Sample Selection} 

The sample galaxies were chosen to be advanced mergers of
galaxies which have already lost their identities and merged
into a single nucleus (up to the resolution limit of 2" in
2MASS), and which still looked disturbed and had signs of
recent interaction like tidal tails, plumes etc, for more
details see Chitre \& Jog (2002). The galaxies were chosen from
the Atlas of peculiar galaxies (Arp 1966) and the Arp-Madore
catalogue of southern peculiar galaxies (Arp \& Madore 1987).

Our earlier work (Chitre \& Jog 2002) showed that these could be
divided into three dynamically distinct classes: where the
luminosity profiles could either be fitted by an $r^{1/4}$
profile (class I), or an outer exponential (class II), or a
no-fit (class III). In the present paper, we have chosen representative
examples from each of these classes, with an emphasis on class
III which showed sloshing of the inner isophotes even on a
visual inspection (Chitre \& Jog 2002). Of that sample of 27
galaxies, only those galaxies which had a large enough angular
size for non-axisymmetric analysis were chosen. This meant
that they should be divisible into  at least 5-6 annular rings,
with each ring size about twice the resolution of the 2MASS data,
so that a meaningful non-axisymmetric analysis could be carried out.
The J-band data was used for this purpose for all the galaxies 
since the images are deeper in this band allowing for a radial sampling.

This led to a total of eight galaxies to be selected. To this, we
added a further four more galaxies from the Arp and Arp-Madore catalogs
for which such a non-axisymmetric analysis could be carried out,
to supplement our sample: namely, Arp 209, Arp 254,
AM 1025-433, and AM 0612-373.

Thus, the total sample used for the present work consists of 12 galaxies
divided into the three classes as follows:

\noindent Class I : Arp 221, Arp 222, Arp 225, AM 0612-373  

\noindent Class II: Arp 162, Arp 212, Arp 224 

\noindent Class III:
Arp 160, Arp 163,  Arp 209, Arp 254, AM 1025-433.

The four new galaxies chosen here which were not in our earlier sample
fall into class I (AM 0612-373) and class III (Arp 209, Arp 254, 
AM 1025-433): the luminosity profiles defining the
classification of all the galaxies are given below.

For the radial fits, and the isophotal analysis we used the K$_s$
- band data for all galaxies. This is because the dust
extinction effects are least important for this band (than
say J or H) and hence this is suitable for the central sloshing
to be studied via the isophotal analysis.

 The 2MASS full-resolution images for extended sources were obtained from 
 the 3-dimensional FITS image cubes or from the mosaic images
for larger galaxies. The  K$_s$ band images were analysed using the 
task ELLIPSE within STSDAS\footnote{STSDAS is a product of 
the Space Telescope Science Institute, which is operated by AURA for NASA.}. 
 The procedure consisted of fitting elliptical isophotes to the galaxy 
 images and deriving the one-dimensional azimuthally averaged radial profiles 
 for the surface brightness, ellipticity, position angle, etc. based on the 
 algorithm given by Jedrzejewski (1987). The centre, position angle and 
 ellipticity were allowed to vary. Ellipses were fit right up to the 
 central pixel. The parameters thus derived, namely the surface brightness 
 profile, the ellipticity and the position angle variation, 
 and x0, y0 -the x and y coordinates of the centres, were studied.

\begin{figure*}
{\resizebox{18cm}{18cm}{\includegraphics{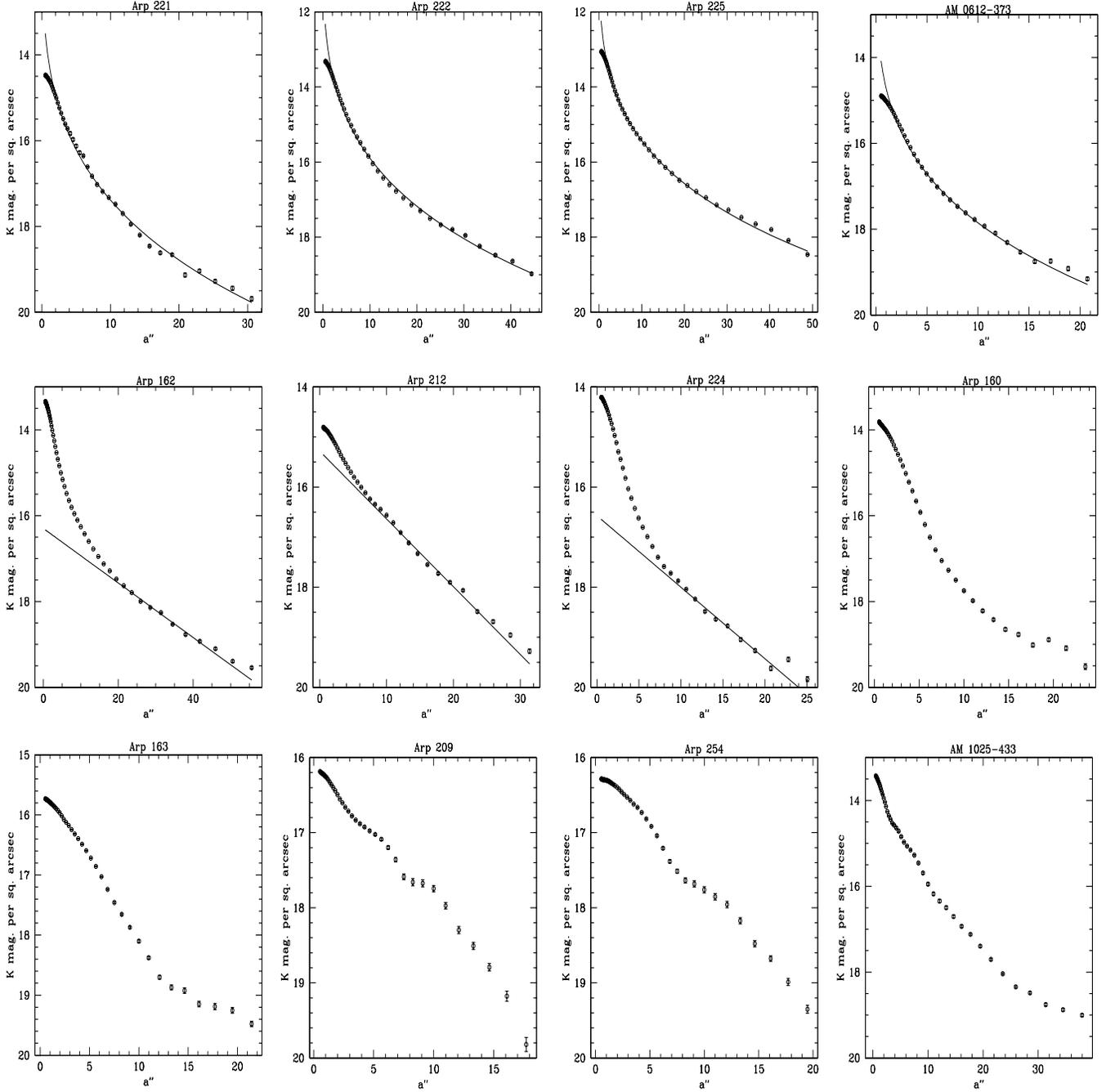}}}
\caption{The K$_s$-band magnitude in mag arc sec$^{-2}$ is plotted 
against the semi-major axis, a" in arc sec. The first four galaxies (Arp 221, Arp
222, Arp 225, and AM 0612-373)  are well fit 
by an r$^{1/4}$ law (class I); the next three (Arp 162, Arp 212,
and Arp 224) are well fit by an outer exponential (class II), and the 
last five are no-fit galaxies (class III).}
\end{figure*}

The resulting surface
brightness radial profiles are plotted for all the 12 sample
galaxies in Figure 1.   The luminosity profile-fitting
for eight of these galaxies from our earlier sample has already been done 
(see Fig.1 in Chitre \& Jog 2002). As discussed in that paper,
surprisingly, the profiles can be best fit by an outer
exponential in four cases; and the fit is robust in the sense
that the radial range over which the fit is valid is typically 2
times the disc scalelength or larger, and the error bars are small.
The values of the de
Vaucouleurs  scalelength, R$_e$ (for class I galaxies), and
the exponential disc scalelength, R$_D$ (for class II galaxies)
are given in Table 1.  This figure defines the 
classification for the four new galaxies as follows: AM 0612-373 (class I),
 Arp 209, Arp 254, and AM 1025-433 (class III).

\begin{table}
\centering
  \begin{minipage}{140mm}
   \caption{Basic data on mergers of galaxies}
\begin{tabular}{lllll}
Name& Alt. name & $R_e$ , $R_D$&velocity& 1$\arcsec$%
\footnote{Using $H_0$ = 75 $km s^{-1} Mpc ^{-1}$}\\

&&(kpc)&(km s$^{-1}$)&(pc)\\
Class I\\
Arp 221 & -&5.76$\pm$0.41&5546&358\\
Arp 222 &NGC 7727&2.83$\pm$0.46& 1855 & 125\\
Arp 225 &NGC 2655&3.26$\pm$0.38& 1404& 91\\
AM 0612-373&-&11.26$\pm$0.45 & 9864 & 620\\
&&&&\\
Class II\\
Arp 162 &NGC 3414&1.56$\pm$0.04& 1414 & 87\\
Arp 212 &NGC 7625&0.85$\pm$0.03& 1633 & 104 \\
Arp 224 &NGC 3921&2.91$\pm$0.21& 5838 & 383 \\
&&&&\\
Class III\\
Arp 160 &NGC 4194&-& 2442 & 168\\
Arp 163 &NGC 4670&-& 1069 & 78\\
Arp 209&NGC 6052&-& 4716 & 305\\
Arp 254 &NGC 5917&-& 1904 & 123\\
AM 1025-433&NGC 3256 &-& 2814 & 182\\
\end{tabular}
\end{minipage}
\end{table}

The Table 1 lists the sample chosen, distributed in terms of this
classification, with all the relevant astronomical parameters
(Galaxy name, Alternate name, R$_e$,R$_D$,  velocity, and 1" in pc) 
for the three classes showing the different radial fits.

\begin{table}
\centering
 \begin{minipage}{140mm}
  \caption{Average central asymmetry properties for mergers}
  \begin{tabular}{llll}
Name & Sloshing of centre & $<A1>$ & $<A2>$\\
& within 1 kpc& within 5 kpc & within 5 kpc\\
Class I\\
Arp 221 & 0.08 & 0.16 & 0.09\\
Arp 222 & 0.10 & 0.04 & 0.18\\
Arp 225 & 0.09 & 0.03 & 0.29\\
AM 0612-373 & 0.10 & 0.05 & 0.17\\
{\bf Avg.: Class I} & 0.09 & 0.07 & 0.18 \\
&&&\\
Class II\\
Arp 162 & 0.06 & 0.02 & 0.24\\
Arp 212 & 0.08 & 0.11 & 0.19\\
Arp 224 & 0.14 & 0.23 & 0.16\\
{\bf Avg.: Class II} &  0.09 & 0.12 & 0.20 \\
&&&\\
Class III\\
Arp 160 & 0.27 & 0.15 & 0.30\\
Arp 163 & 0.31 & 0.15 & 0.19\\
Arp 209 & 0.04 & 0.23 & 0.22\\ 
Arp 254 & 0.06 & 0.09 & 0.34\\
AM 1025-433 & 0.20 & 0.16 & 0.29\\
{\bf Avg.: Class III} &  0.18 & 0.16  & 0.27 \\
\end{tabular}
\end{minipage}
\end{table}

\section{Data Analysis and Results}

\subsection{Isophotal Analysis: Sloshing}

The isophotal analysis of the images is carried out as described
in the last section. The centre, position angle and 
 ellipticity were allowed to vary to obtain the best fit by
ellipses to the galaxy image.

 In Figure 2, we plot a block of 4
diagrams for each galaxy, which shows the centres of the
isophotes (x0, y0 vs. $a$, the semi-major axis of the elliptical
isophote (in units of kpc). Also shown are the ellipticity and
the position angles (PA) versus $a$, the semi-major axis. Clearly,
in most cases, the centres of the isophotes do not remain
constant. Instead, they show a wandering or a sloshing behaviour.
{\it This indicates a dynamically unrelaxed central mass
distribution}. In general, the sloshing is higher in the class
III galaxies which are also unrelaxed in the outer regions.
This is confirmed by a quantitative measurement of normalized 
sloshing within $a=$1 kpc, which is
 given as average of $(((x-x0)^2+(y-y0)^2))^{1/2}/a$ - see
Table 2. This allows a comparison of sloshing seen in the
various classes of galaxies.
\begin{figure*}
{\resizebox{15cm}{22cm}{\includegraphics{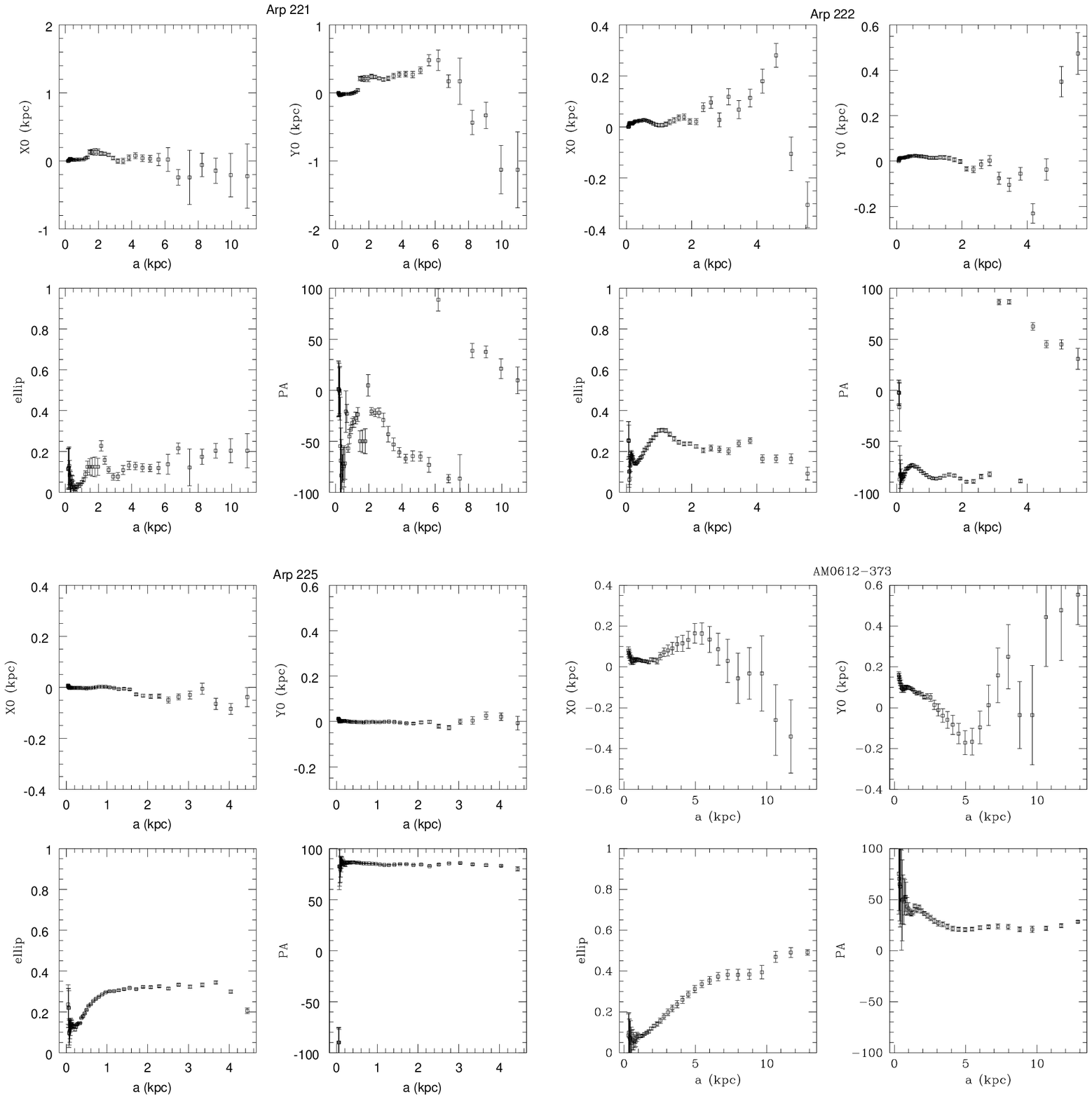}}}
\caption{Plots of isophotal analysis for the K$_s$-band data for the 
sample galaxies. 
For each case, we plot a block of 4 plots, showing the centre $x0, y0$ 
vs the semi-major axis $a$  of the fitted isophotes, and also the 
ellipticity and the position angle (PA) versus $a$. All the
galaxies on this page are class I galaxies.}
\end{figure*}
\begin{figure*}
{\resizebox{15cm}{22cm}{\includegraphics{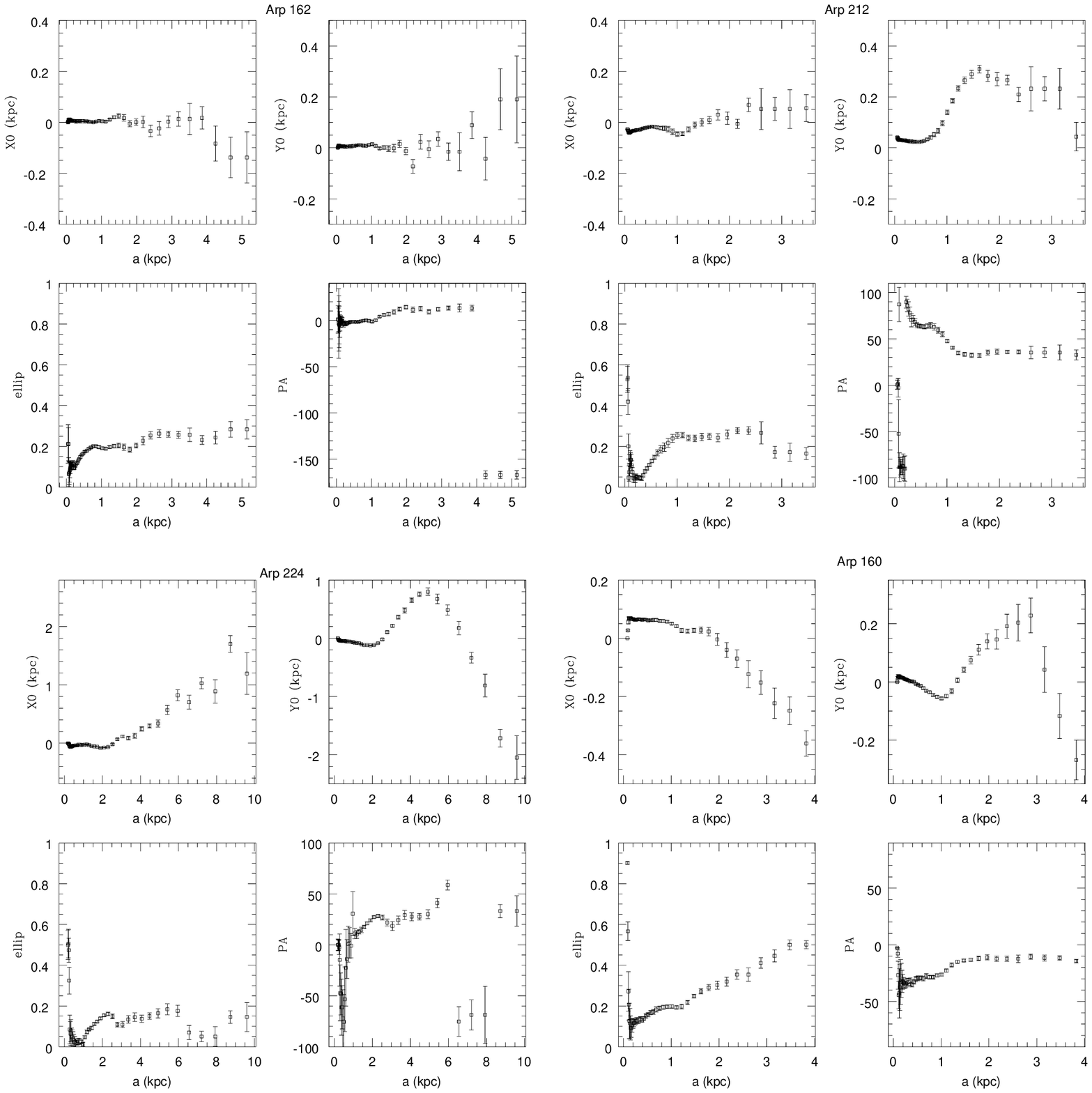}}}
Fig 2(contd.)

Arp 162, Arp 212, and Arp 224 are
class II type galaxies, while Arp 160 is a class III type galaxy- it
shows a larger sloshing of the centre (x0, y0) of isophotes.

\end{figure*}
\begin{figure*}

{\resizebox{15cm}{22cm}{\includegraphics{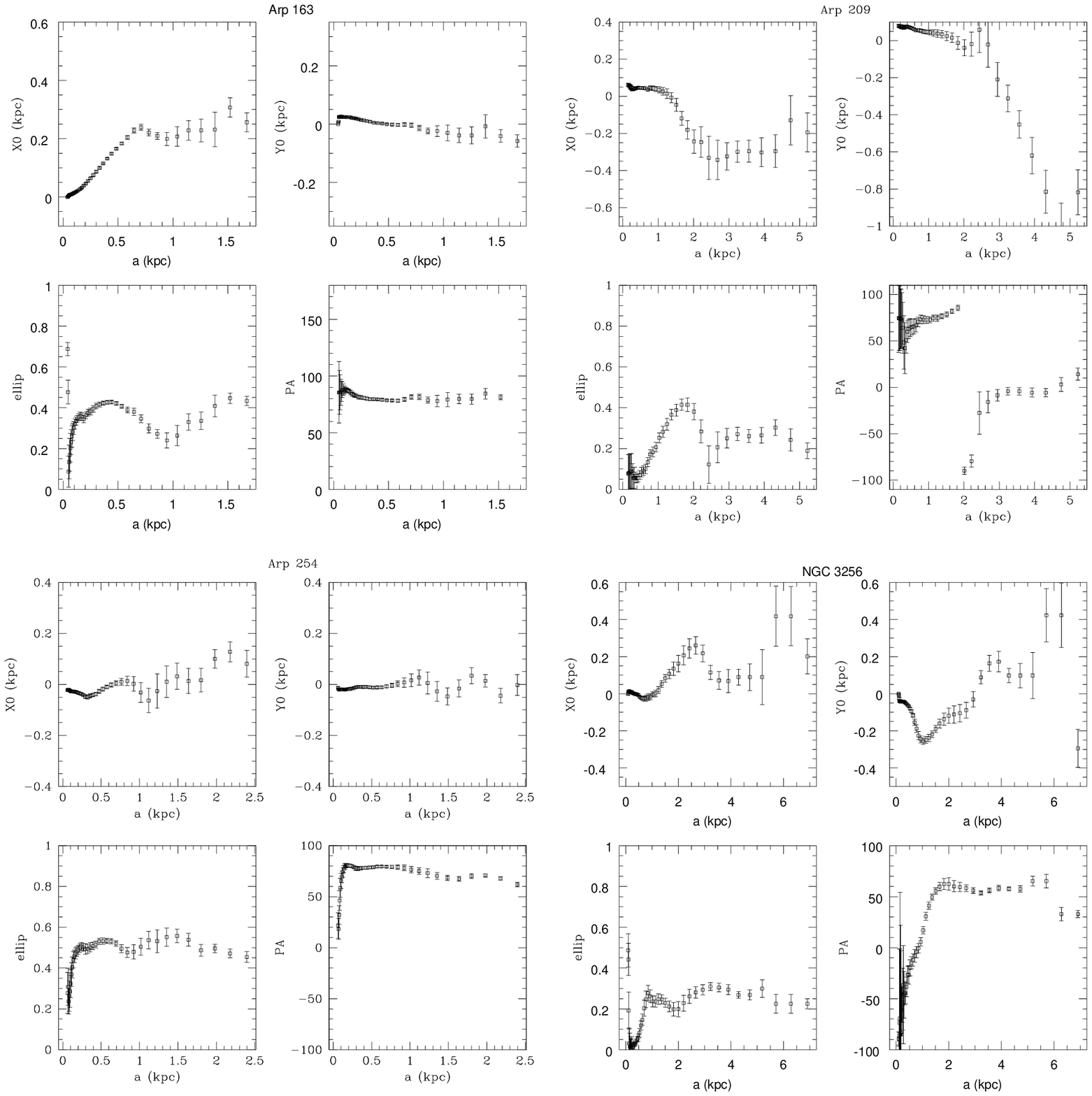}}}
Fig 2(contd.)

All the galaxies shown are class III type- and they show larger
sloshing than shown by Class I or II.

\end{figure*}

\subsection{Non-axisymmetry: Fourier Amplitudes}

The images from 2MASS of mergers were Fourier-analyzed to obtain the
amplitudes and phases of the
Fourier components m = 1, 2,3 and 4, measured w.r.t. a constant centre.
A similar approach has been used in the past to
divide the near-IR images of normal galaxies into radial bins,
and measure the deviation from axisymmetry in terms of the
fractional Fourier amplitudes such as A$_1$ and A$_2$ measured
w.r.t. A$_0$ the average amplitude,
where A$_1$ denotes the amplitude for lopsidedness (m=1) and A$_2$
denotes the amplitude for bars or disk ellipticity or spirals
(e.g., Rix \& Zaritsky 1995, Laurikainen et al. 2003, Buta et al. 2005).

In our analysis, the centre
is kept constant and the Fourier analysis is carried out over
annular rings w.r.t. this centre. For this, we sum up the
pixel values in the radial bins so that the complete 2-D data is
used and we do not try to fit a curve. This procedure is similar
to the one used by Rix \& Zaritsky (1995).
 The resulting A$_1$ and A$_2$
amplitudes for m=1 and 2 are better indicators of mass asymmetry measured
w.r.t. the constant centre (than the usual boxiness etc for an isophote), 
especially since 
the centres of isophotes show a lot of sloshing in the case of mergers.
For example, it is easy to see that if we use the standard isophotal analysis,
 then the azimuthal average taken over an
isophotal contour w.r.t. the
isophotal centre would give a zero average
lopsidedness since the average of cos $\phi$ would be zero.

Thus, for measuring the non-axisymmetric mass
distribution we need to keep the centre constant and then
measure the Fourier amplitudes over annular radial bins, for this a special procedure was
adopted which is described below.

First, the  x and y co-ordinates of the centre of the galaxy 
(or the stellar centre) were determined using 
the task "imexam" in IRAF and plotting the radial profile.
Once this pixel value was found in terms 
    of the x and y co-ordinates of the frame, we designated the rest of 
    the image pixels 
    in terms of the distance $r$ and the angle $\theta$ from the centre.
The image was then cut into annular regions. A polar coordinate grid was
centered on the 
    galaxy nucleus with 36 azimuthal bins with 10 degrees per bin, 
    and  the number of radial bins depended on the size of the galaxy. 
In each case, we took the radial width of the annular ring to be at least 2 
times the resolution of the image.  
The pixel intensity values in each bin were summed to give
an average value per bin. Using the task
"nfit1d" within STSDAS, we fit the following function, $f$ to each
annular region:

$$ f = A_0 + A_0 * [ A_1 * cos (x - p_1) + A_2 * cos (2 x - 2
\times p_2)$$
$$ \: \: \: \: \: \: \: \: \: \: \: \: \: \: \: + A_3 * cos (3 x
 - 3 \times p_3)
+ A_4 * cos (4 x - 4 \times p_4) ] \eqno (1) $$

\noindent where the $A_1, A_2$ etc denote the fractional Fourier
amplitudes and the $p_1,p_2$ etc denote
the phases of the various Fourier components.
Each annular region in the image gives one set of values for the
above fitting. The number of points we get in the final plot
are the same as the number of annular regions that were made in the 
original image.

\begin{figure*}
{\resizebox{15cm}{22cm}{\includegraphics{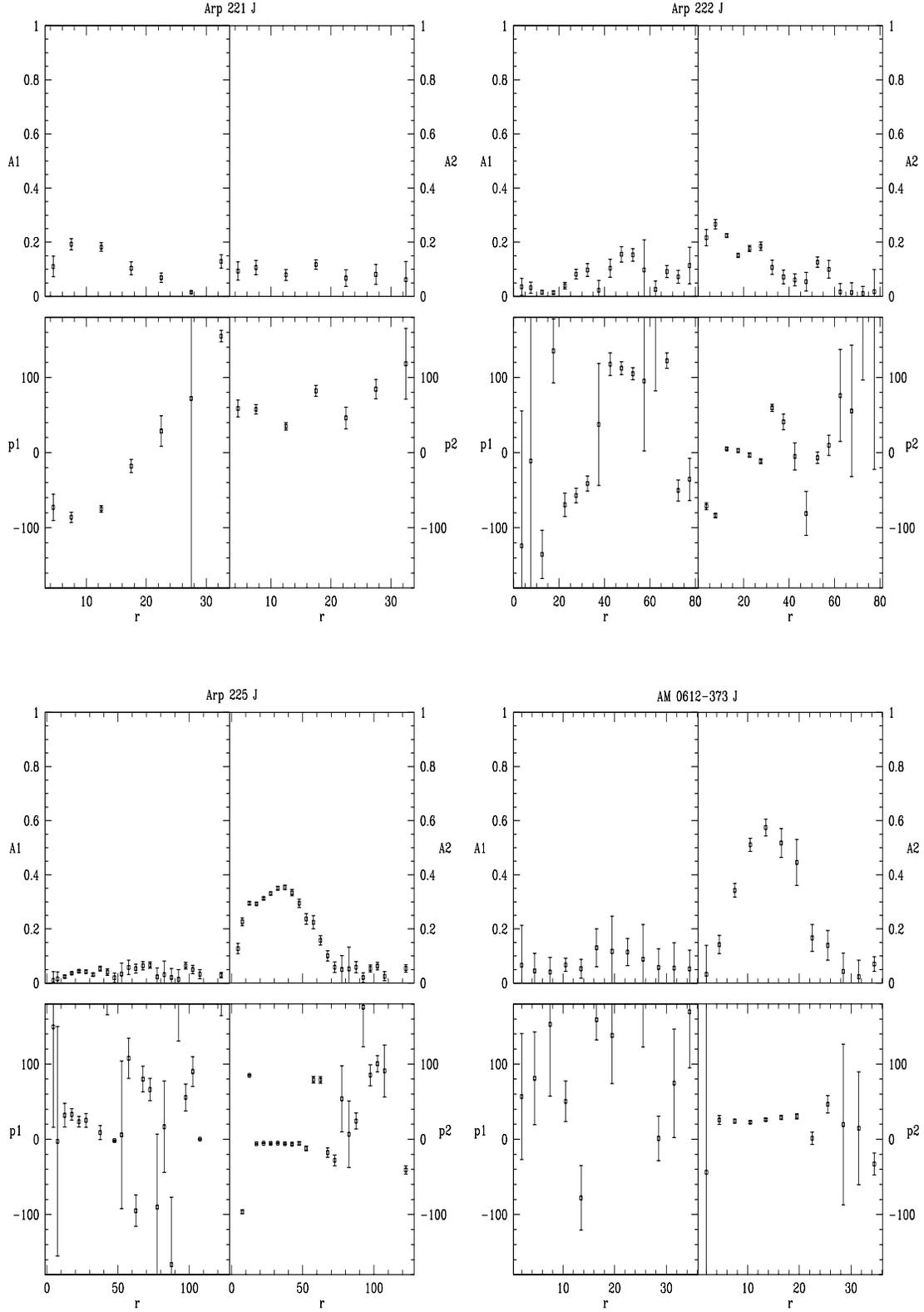}}}
\caption{Plots of Fourier amplitudes and phases A$_1$ and P$_1$ and A$_2$ 
and P$_2$ versus R, obtained from a  non-axisymmetric analysis with a 
constant centre, for the J-band data for the entire sample studied. All 
mergers, especially the Class III galaxies, show a large amplitude
of lopsidedness A$_1$.}
\end{figure*}
\begin{figure*}
{\resizebox{15cm}{22cm}{\includegraphics{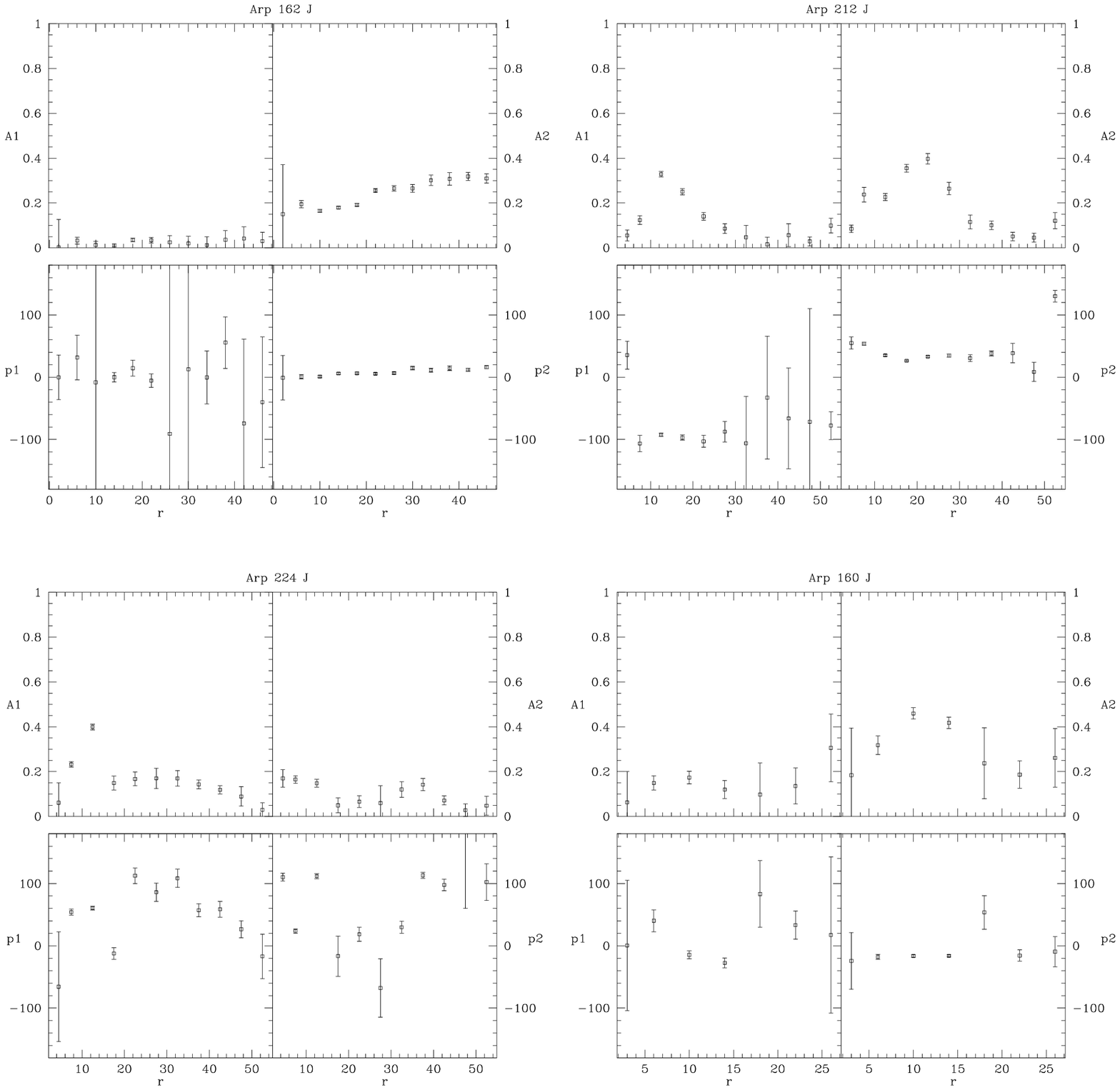}}}
Fig 3(contd.)
\end{figure*}
\begin{figure*}
{\resizebox{15cm}{22cm}{\includegraphics{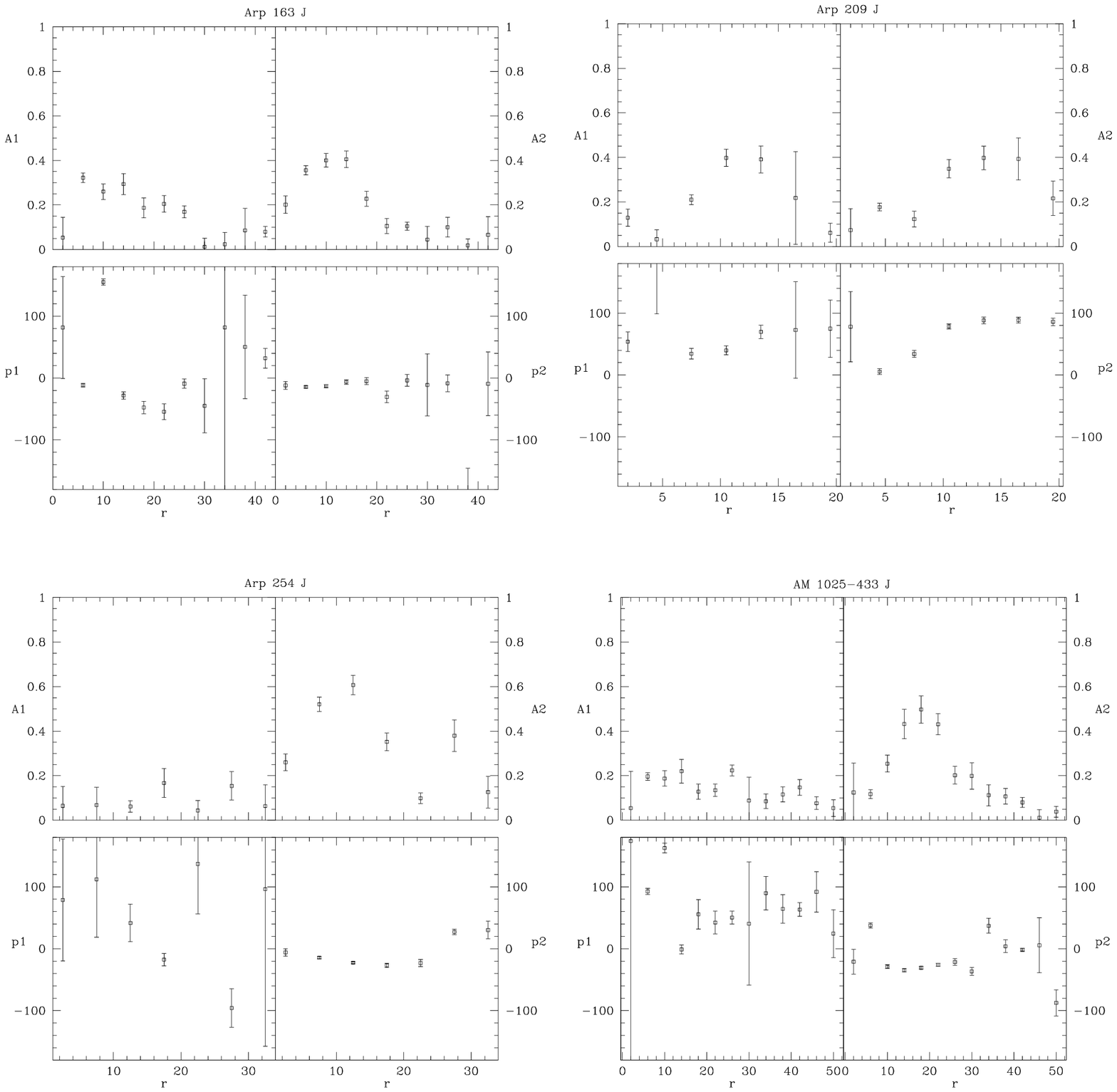}}}
Fig 3(contd.)
\end{figure*}

 In Figure 3, we plot the fractional amplitudes $A_1, A_2$  and
the phases $p_1, p_2$ versus radius $r$ (in $\arcsec$) for
m=1 and 2 respectively. The m=3 and 4 amplitudes are generally smaller, and
m=3 is important only when m=1 is large. For the sake of
brevity, we do not show the amplitudes and phases for m=3 and 4
for the sample. However, in the notes on individual galaxies
(Appendix A) we also add comments about A$_3$ and A$_4$ when relevant.

Figure 3 shows that all the mergers studied show a high value of lopsidedness
(A$_1$), the values are generally higher for the class III
galaxies - see Table 2 for the detailed values including the
averages over the classes. The average values for A$_1$ and
A$_2$ are given over the central 5 kpc in each merger since it represents a 
typical central region, and also this allows a comparison
between different galaxies.

In order to avoid any spurious variation with class introduced due to the
dependence on the band chosen, we have taken homogeneous data (in
J-band from 2MASS) for all the galaxies for the above
non-axisymmetric analysis (Fig. 3), and the K$_s$-band data from
2MASS for the
isophotal analysis (Fig. 2).

Figure 3 shows that the values of A$_2$ in mergers are also large.
Normally, the values of A$_2$ are  taken to denote bars or
spirals or disc ellipticity
(Rix \& Zaritsky 1995, Buta et al. 2005, Bournaud et al. 2005 a). 
It is hard to see how spiral features can survive the strong disturbance
in a merger.
We add the evidence from the isophotal analysis and find that the 
characteristic feature of bars, namely increasing ellipticity and a nearly 
constant position angle (PA) (e.g. Wozniak et al. 1995) is seen in only 
three cases, which we conclude  have well-defined bars. 
These are:   Arp 160, Arp 162, and Arp 163.

Grosbol et al. (2004) state that a circular disc will appear elliptical 
when projected and this could give rise to an artificial bisymmetric 
component with a constant phase. This could be used to explain the high 
values of A$_2$ seen in many galaxies in our sample. In fact
this effect is seen in all the galaxies except Arp 221 (Class
I), Arp 212 (Class II) , and Arp 209 (Class III). Thus, in most
cases studied, the merger remnants seem to indicate a preferred disc plane.
This is seen even in three of the class I galaxies with 
an outer elliptical-like profile (Arp 222, Arp 225, AM0612-373),
which is unexpected.

We note, however, that the previous work on 
deriving the Fourier coefficients to get the asymmetry has been done for 
spiral galaxies. These galaxies were deprojected by estimating their 
inclinations assuming that the discs are intrinsically round. It is 
reasonable to talk about projection effects in the case of spirals and also 
comparatively easy to deproject these galaxies. However, here we are 
looking at systems completely or partially distorted by the interaction or a 
merger and hence it is not possible to estimate the inclination of the disc 
component and correct for inclination effects. A further discussion on the 
variation of Fourier amplitudes with
class types in our sample is given in Section 4.2.

\subsection{Comparison with non-merger galaxies}

So far we have studied the mergers and we have shown these to have
significant central non-axisymmetry. In order to check that the origin
of this is
related to the merger history, we next carry out a similar
analysis for a control sample of non-merging, normal spiral galaxies and
show that these indeed have lower central asymmetry.

The selection of a sample of non-merger galaxies is  
by no means a trivial issue.
We do not include any early-type galaxies which are now believed to evolve 
from late-type galaxies via secular evolution or galaxy mergers. The
galaxies selected are nearly face-on (for easier analysis), and
are late-type and not strongly barred and not active - so as to
increase the chance of their having had no interactions in the recent past. 
The set of eight such normal or non-merger galaxies which are also large
enough in angular size to allow the non-axisymmetric analysis in
J-band (see Section 2) were chosen from
the 2MASS Large Galaxy Catalog (Jarrett et al. 2003), and these are:
NGC 628 (M74), NGC 3147, NGC 4254 (M100),  NGC 4321 (M99), NGC 4540, 
NGC 4689, NGC 5248, NGC 5377.

These show an outer exponential fit as expected for normal
spiral galaxies.
The basic data for these, including the derived disc scalelength (R$_D$), 
is listed in Table 3.

\begin{table}
\centering
  \begin{minipage}{140mm}
   \caption{Basic data on normal galaxies}
\begin{tabular}{lll}
Name& Alt. name& ${R_D}\arcsec$%
\footnote{Using $H_0$ = 75 $km s^{-1} Mpc ^{-1}$}\\
&&(kpc)\\
NGC 628 & M74 & 2.51$\pm$0.13 \\
NGC 3147 & - & 3.82$\pm$0.09 \\
NGC 4254 & M100 & 9.48$\pm$0.75 \\
NGC 4321 & M99 & 3.81$\pm$0.27 \\
NGC 4540 & - & 1.37$\pm$0.04  \\
NGC 4689 & - & 4.39$\pm$0.32  \\
NGC 5248 & - & 7.34$\pm$1.79  \\
NGC 5377 & - & 3.70$\pm$0.21  \\
\end{tabular}
\end{minipage}
\end{table}

\begin{table}
\centering
 \begin{minipage}{140mm}
  \caption{Average central asymmetry for normal galaxies}
  \begin{tabular}{llll}
Name   & Sloshing of centre & $<A1>$ & $<A2>$\\
& within 1 kpc& within 5 kpc & within 5 kpc\\
NGC 628 &  0.07 & 0.04 & 0.10\\
NGC 3147 & 0.07  & 0.03 & 0.08\\
NGC 4254 & 0.09  & 0.02  & 0.14 \\
NGC 4321 & 0.02  & 0.03 & 0.15 \\
NGC 4540 & 0.13  & 0.10 & 0.15 \\
NGC 4689 & 0.10  & 0.06  & 0.10 \\
NGC 5248 & 0.06  & 0.07  & 0.34 \\
NGC 5377 & 0.04 & 0.02    & 0.53 \\
{\bf Average:}   & 0.07 &  0.04 &  0.20\\

\end{tabular}
\end{minipage}
\end{table}

The isophotal analysis for the K$_s$-band data (as in Section 3.1) and 
the Fourier analysis for the J-band data
(as in Section 3.2) were carried out for this sample. The
resulting values of sloshing within 1 kpc and the mean A$_1$ and
A$_2$ within 5 kpc are given in Table 4. We find that these show
a lower magnitude for the sloshing of isophotes and a 
lower mean central A$_1$ value (compare with Table 2).
 The non-merger or normal galaxies show sloshing values which are smaller by a
factor of 2 compared to the un-relaxed (or class III) mergers. The
Fourier amplitudes for lopsidedness for the normal galaxies are smaller 
by a factor of
2-3 compared to class I and II remnants and smaller by a factor
of 4 compared to the class III remnants.
Thus, the origin for the high central asymmetry (especially
A$_1$) observed in the merger remants
can be attributed to its merger history.

Interestingly, the mean A$_2$ value for this sample is
comparable to the class I and II galaxies, thus providing
additional support for our interpretation of this (Section 3.2) as being due
to an inclined disc. The higher values for A$_2$ in the
unrelaxed (class III) type galaxies then could be due to a
thicker disc seen at an angle.


\section{Dynamical Implications}

\subsection{Long-lived Central Asymmetry:}

The one clear result from this work is that all mergers show
high levels of asymmetry in their central few-kpc regions,
even in systems that have relaxed in the outer parts. The
large sample of galaxies used here gives a
statistical idea of the in-situ non-axisymmetry in  merger remnants
with a variety of dynamical properties.
N-body simulations of unequal-mass mergers by Bournaud et al.
(2004) have shown that the advanced remnants with signs of
tidal interactions and disturbed profiles in the outer regions
(beyond say 20 kpc or so) have ages  $\leq 2$ Gyr. 
 Class I and II represent
mergers of different mass ratios, of masses 1:1-3:1, and 4.5-10:1
respectively (Bournaud et al. 2005 b), while Class III could be 
 younger remnants ($\sim 1$ Gyr). The dynamical timescale
in the central few kpc where the random motion are $\sim$ a few 100 
km s$^{-1}$ (e.g., Jog \& Chitre 2002) is $\sim 10^7$ yr.
 Thus the central regions in mergers are not relaxed in over a
few hundred local dynamical timescales.  This is an important and a robust
dynamical result, and has important implications for the
evolution of the very central regions of mergers. 

For example, such asymmetries could play an important role in
the fueling of the central black-hole in an AGN since they
provide a means of outward transfer of angular momentum.
The central asymmetries can help fuel a central black hole for a long
time $\sim$ 100 dynamical timescales. Also these can be
important in the gradual build-up of bulges - on lines of
bars as an intermediate stage in the build-up of boxy bulges by mergers 
(see e.g. Walker et al. 1996).

It is instructive to compare the central lopsidedness in mergers with the
lopsidedness in the outer discs of normal isolated spiral
galaxies. The latter was initially believed to be
short-lived ($< 1 $ Gyr) and to get wound up in a few dynamical timescales
(Baldwin et al. 1980). However, more recent N-body studies
of evolution of disc lopsidedness have shown that these features
can be surprisingly long-lived to more than 2 Gyr or several
tens of disc dynamical timescales, although the physical reason for
this is still not understood (Bournaud et al. 2005 a).
The results from the present paper confirm the asymmetry in the
central regions to be also long-lived and the ratio of its 
lifetime to the local dyanmical timescales is even 
higher $\sim 100$.

It should be noted that central regions of
normal galaxies like our Galaxy, M 31 and M 33 also show sloshing (Miller \&
Smith 1992, also see Section 3.2) where the centre is off-set with respect to the outer isophotes
but the other isophotes are concentric and the off-set is
believed to be due to overstability,  whereas in the mergers we
have studied, the isophotes are steadily off-centered with respect to each
other. Thus the origin and evolution of central asymmetry could
be different in these cases, which is not surprising given the
totally different systems being studied.

\subsection {Lopsidedness and bars in Mergers:}

{\bf 1. Lopsidedness in centres of mergers:}  The amplitude of $m=1$ indicates 
lopsidedness and is not
affected by inclination effects (e.g., Rix \& Zaritsky 1995). 
This is particularly important in the study of mergers where the disc
plane may not be so well-defined in the central regions.

We note that the physical parameters A$_1$, the amplitude and p$_1$, 
the phase of lopsidedness show a  different behaviour in mergers 
than that in the outer parts of normal galaxies. In the
centres of mergers the amplitude A$_1$ peaks at an intermediate 
radius of a few kpc and then decreases at larger radii (see
Fig. 3), and the phase p$_1$ shows large fluctuations with radius - see for
example the results for Arp 221, Arp 224 and Arp 163 in Table 3.
In contrast, in the outer parts (beyond $\sim 1.5$ disc scalelengths
or $\sim $ 5 kpc) of normal galaxies the A$_1$ values show a steady increase
with radius, and the phase p$_1$ is nearly constant with radius -
as seen from
Rix \& Zaritsky (1995), and  Angiras et al. (2006). This
constancy in phase indicates a global distortion (Jog 1997). 
Thus these two features clearly point to a different physical
origin and evolution of lopsidedness in centres of mergers.

Our sample is chosen
so that the two nuclei have coalesced to $<$ 2 " or the
resolution of the data, or an average separation of $<$ 500
pc (see Table 2). The dynamical
evolution of the central regions of mergers including the long lifetime of
the central lopsidedness needs to be studied theoretically.

\medskip

\noindent {\bf 2. Interpretation of A$_2$ values:} The high values of A$_2$ measured in the merger
remnants here is unexpected, since a merger 
is expected to
result in the disruption of a bar (Pfenniger 1991, Athanassoula \&
Bosma 1999). 
Also, these bars are stronger compared to the bars in normal galaxies
(compare Table 2 in our paper with the Appendix in Bournaud et al. 2005 a). 

It is interesting that in two of our sample galaxies which show a
bar (Arp 160 and Arp 163), the outer regions are not
relaxed and hence cannot be fit by either an r$^{1/4}$ or an
exponential profile, and yet the central region show a strong
bar and lopsidedness. Thus the orbit building for the bars seems
to proceed irrespective of the outer disturbed regions.  
The other galaxy showing a bar, Arp 162,
is a class II galaxy which has a relaxed, exponential outer profile.

The high values of A$_2$ measured in the other, non-barred galaxies could 
be due to an
inclined thick disc, as discussed in Section 3.2. The class II galaxies are 
hybrid systems
with high random velocities and thick discs (Jog \& Chitre 2002)
and these have been shown to arise naturally in mergers of
unequal-mass galaxies covering a range of 4.5:1-10:1 (Bournaud
et al. 2004). The high values of A$_2$ seen in the sample
mergers of class II can thus be naturally  explained as arising 
due to an inclined thick disc. 

The values of A$_2$ measured are higher for all class III galaxies.
The high values of A$_2$ if taken as an indication of an
inclined thick disc- in no-fit galaxies, which do not have a
bar (such as Arp 254 and AM 1025-433),  indicates 
 the existence of a galactic plane even if the outer regions are
highly disturbed.

The high A$_2$ values seen in Class I galaxies with
elliptical-like profiles is even more surprising. The
isophotal analysis shows that these do not show the pattern expected of a 
bar, and we have argued that this could be due to an
inclined disc (Section 3.2), thus these seem to be unusual systems.

In any case, it is clear that the central
asymmetries (m=1 and m=2) seen in class I  remnants indicate
that they are dynamically still far from being regular elliptical
galaxies. This is extra evidence in addition to the
other arguments such as the high gas content, extended tails and outer 
disturbed profiles, and different kinematics seen in mergers
which have been used to show that merger remnants and ellipticals are not identical 
(e.g., Kormendy \& Sanders 1992, Shier \& Fisher 1998).

\medskip

\noindent {\bf 3. Variation with mass ratios and epoch
of mergers:}
There is no discernible difference in the sloshing values or the 
A$_2$ values for class I and II (see Table 2), while the A$_1$
values are  higher for class II galaxies. Thus the mass ratio
of the progenitor or merging galaxies does not seem to significantly 
affect the degree 
of central non-axisymmetry in mergers at a later stage in
its evolution, when the outer parts have relaxed into an
elliptical-like (class I) or an exponential (class II) luminosity profile.

However, the stage of relaxation does seem to be a significant
factor, because the class III galaxies, which are in early stages of 
relaxation do show higher amplitudes of sloshing and lopsidedness.
 The various galaxies can thus be arranged in a time
sequence for variation of central asymmetry as follows: the
normal or non-merger galaxies (lowest asymmetry), young merger remnants
of $< 1 Gyr$ or class III mergers (highest asymmetry), evolved merger remnants of $<
2 Gyr$ with
relaxed outer regions or class I and II mergers (higher asymmetry compared to normal galaxies but lower
compared to the young remnants). The dynamical evolution of the 
non-axisymmetric amplitude
 and its relation to the relaxation of the outer 
regions, need to be studied by future N-body simulations.

\subsection {Sloshing values and Fourier amplitudes:}

While both sloshing and the Fourier amplitudes represent asymmetry,
they do not represent the same physical quantity. For example, a galaxy
which has concentric isophotes and hence shows no sloshing 
could still have a non-axisymmetric mass distribution as in a
bar or a spiral arm (m=2). Thus the sloshing represents odd values
of non-axisymmetric Fourier amplitudes. Indeed the
typical sloshing and A$_1$ values (see Figs. 2 and 3, or Table
2) are correlated - although it must be kept in mind that the
sloshing represents only the inner regions ($<$ 1 kpc) of a galaxy
while the non-axisymmetric analysis is typically carried out over
a larger radial range of up to 5 kpc or more.

On comparing with the values for the normal galaxies (Tables 2
and 4), we find that these have comparable values for sloshing
(especially the class I and II mergers, and the normal galaxies sample), but
the A$_1$ values for the mergers are distinctly higher than for
the normal galaxies.

\section{Summary}

We have measured the non-axisymmetry in the mass distribution within
the central few kpc of advanced mergers of galaxies which have
merged into a single nucleus but have indications of
interactions including tidal tails.  The main results obtained are:

\noindent {\bf 1.} The mergers show strong non-axisymmetry - with
the centres of isophotes showing a sloshing by $\sim 20-30 \%$
within the central 1 kpc. The asymmetry is also 
high, as measured by the Fourier amplitudes of the central light
distribution.
The typical fractional lopsided amplitude (A$_1$) within the central 5 kps
is found to be high
$\sim$ 0.12 going upto 0.2, while
the typical A$_2$ values are higher $\sim 0.2$ going
upto 0.3.  
This implies the presence of bars as in Arp 160, Arp 162, and
Arp 163 or
thick discs as in the rest of the sample, both of which are unexpected in mergers with elliptical
profiles or in no-fit, unrelaxed galaxies.

The corresponding values especially for the lopsidedness for a control sample of non-merger
galaxies are smaller by a factor of 2-4, this confirms that the high central
asymmetry in mergers discovered and measured in this paper 
 can be truly attributed to the merger history.

\noindent {\bf 2.}
The ratio of masses of galaxies undergoing the
merger does not have a strong influence on the value of
the central asymmetry once the outer regions have relaxed
(Sections 1 and 4.2).

\noindent {\bf 3.} The stage of merger, however, does seem to be 
significant  because the galaxies where the outer regions are non-relaxed, 
show a higher amplitude of asymmetry in the inner regions
(Section 4.2).

\noindent {\bf 4.} The mergers are about 1-2
Gyr old  as shown by the N-body
simulations (Bournaud et al 2004).
Thus in all cases studied, the central asymmetry 
appears to be long-lived, lasting for over 100 local dynamical timescales. 
This can be important for the dynamical evolution of the central regions
of mergers.

\bigskip

\noindent {\bf Acknowledgements:}

We are grateful to the referee, Robert Jedrzejewski, for  
constructive comments and particularly for suggesting that we
add a comparison with a
control sample of non-merger galaxies.

This publication makes use of data products from the Two Micron
All Sky Survey (2MASS), which is a joint project of the
University of Massachusetts and the Infrared Processing and
Analysis Center/California Institute of Technology, funded by the
National Aeronautics and Space Administration and the National
Science Foundation.

\bigskip

\noindent {\bf Appendix A: {Notes on Individual Galaxies:}}

\bigskip

The details of sloshing  as well as the non-axisymmetric Fourier
amplitudes obtained for each galaxy in Section 3 are summarised below:

\medskip

\noindent {Class I galaxies:}

\noindent 1.   Arp 221: The coordinates of the centre 
   remain constant in the inner 4" and change beyond that. The non-axisymmetric
    analysis shows that the A$_2$, A$_3$ and A$_4$ values are small 
    while A$_1$ shows a maximum of 0.25 at 8".  
    
\noindent 2.    Arp 222: The coordinates of the centre
    remain constant in the inner 20" and change beyond that. This galaxy 
    has very low A$_1$ values. A$_2$ maintains a constant value of 0.3 from 
    the  center to about 30". The A$_3$ and A$_4$ values are both very small.
    
\noindent 3.  Arp 225:  The coordinates of the 
    centre are constant in the inner 18". This galaxy shows low values of 
    A$_1$, A$_3$ and A$_4$ but high A$_2$ values. The behaviour of A$_2$ is 
    similar 
    to that seen in Arp 222. A$_2$ maintains a value of 0.3 in the inner 45".
    
\noindent 4. AM 0612-373: 
    There is no substantial change in the coordinates of the center,
with a total change of 1 pixel for x0 and y0. 
    The value of A$_1$ is low, while A$_2$ is quite high, peaking at 0.6 
    at 15".  The A$_3$ and A$_4$ values are fairly high.

\bigskip

\noindent  Class II galaxies:
  
\noindent 5.  Arp 162: The centre co-ordinates x0, and y0 are constant in the 
  inner 14". The values of A$_1$ and A$_3$ are extremely small.
The amplitude A$_2$ starts off at about 0.2  between 6"-20" and then 
increases to 0.3 beyond that, while A$_4$ is low 
  in the   inner 20" and then increases. The phase angle is nearly constant 
  throughout for the m=2 and 4 components.
  
\noindent 6.   Arp 212: The centre co-ordinates x0, and y0 remain
constant in the inner 6" and then change substantially. All A
coefficients show (nearly) double-peaked behaviour. The values
of the A coefficients decrease in the
order: A$_1$ $>$ A$_2$ $>$ A$_3$ $>$ A$_4$.

\noindent 7. Arp 224: The centre  (x0, y0) remains constant in
the inner 6". The value of A$_1$ is high showing a peak of 0.3 at
12". In this case also, A$_1$ $>$ A$_2$ $>$ A$_3$ $>$ A$_4$.

\bigskip

\noindent Class III galaxies:

\noindent 8. Arp 160: This shows a constantly
changing centre  (x0, y0) for the subsequent isophotes. 
The values of  A$_2$ are also high with a peak of  0.45 at 10".

\noindent 9. Arp 163: This also shows a constantly changing centre (x0 and
y0) for the nearby isophotes. The value of the m=2 amplitude
A$_2$  is the strongest, with a maximum of 0.4 between 10"-14".
The value of A$_1$ peaks
at 0.3 at 6", and A$_4$ peaks at 0.3 at 14". The value of  A$_3$ is small
at all radii.

\noindent 10. Arp 209:  The centre co-ordinates  are constant in
the inner 4", and vary thereafter. The values of A$_1$, A$_2$ and A$_3$
 are high.

\noindent 11.  Arp 254: This shows changing co-ordinates  for
the centre. The values of  A$_2$ and A$_4$ are large as compared to 
the values of A$_1$ and A$_3$.

\noindent 12.  AM 1045-433:    This is a Toomre sequence galaxy. Here the
co-ordinates of the centre (x0 and y0) show a
change from the innermost region itself. The value of A$_2$ is high with a peak
value of 0.5 at 20".  This galaxy shows the following values for
the Fourier amplitudes: A$_2$ $>$ A$_3$ $>$ A$_4$ $>$ A$_1$.

\bigskip

\centerline  {\bf  {References}}

\noindent Angiras, R.A., Jog, C.J., Omar, A., \& Dwarakanath,
K.S. 2006, MNRAS, in press (also, astro-ph/0604120)

\noindent Arp, H. 1966, Atlas of Peculiar Galaxies (Pasadena:
California Institute of Technology)

\noindent Arp, H., \& Madore, B.F. 1987, A  Catalogue of
Southern Peculiar Galaxies and Associations (Cambridge:
Cambridge Univ. Press)

\noindent Athanassoula, E. , \& Bosma, A. 2003, Ap\&SS, 284, 491

\noindent Baldwin, J.E., Lynden-Bell, D., \& Sancisi, R. 1980,
 MNRAS, 193, 313

\noindent Barnes, J.E. 1992, ApJ, 393, 484

\noindent Barnes, J.E., \& Sanders, D.B. 1999, "Galaxy
interactions at low and high redshift", IAU symposium 186
(Dordrecht: Kluwer)

\noindent Bournaud, F., Combes, F., \& Jog, C.J.
2004, A \& A, 418, L27

\noindent Bournaud, F., Combes, F., Jog, C.J., \& Puerari, I.
2005 a, A \& A, 438, 507

\noindent Bournaud, F., Jog, C.J., \& Combes, F. 2005 b, A \& A,
437, 69

\noindent Buta, R., Vasylyev, S., Salo, H., \& Laurikainen, E.
   2005, AJ, 130, 506

\noindent Chitre, A., \& Jog, C.J. 2002, A \& A, 388, 407

\noindent Grosbol, P., Patsis, P.A., \& Pompei, E. 2004, A \& A,
423, 849

\noindent Jarrett, T.H., Chester, T., Cutri, R., Schneider,
S.E., \& Huchra, J.P. 2003, AJ, 125, 525

\noindent Jedrzejewski, R.I. 1987, MNRAS, 226, 747

\noindent Jog, C.J. 1997, ApJ, 488, 642

\noindent Jog, C.J., \& Chitre, A. 2002, A \& A, 393, L89

\noindent Kormendy, J., \& Sanders, D.B.. 1992, ApJ, 390, L53

\noindent Laurikainen, E., Salo, H., \& Rautiainen, P.  2002, MNRAS, 331, 880

\noindent Miller, R.H., \& Smith, B.F. 1992, ApJ, 393, 508

\noindent Naab, T., \& Trujillo, I. 2005, MNRAS (submitted), 
astro-ph/0508362.

\noindent Pfenniger, D. 1991, in Dynamics of Disc Galaxies,
Varberg Castle, Sweden. p. 191.

\noindent Rix, H.-W., \& Zaritsky, D. 1995, ApJ, 447, 82

\noindent Schweizer, F. 1982, ApJ, 252, 455

\noindent Schweizer, F. 1996, AJ, 111, 109

\noindent Shier, L.M., \& Fischer, J. 1998, ApJ, 497, 163

\noindent Stanford, S.A., \& Bushouse, H.A. 1991, ApJ, 371, 92

\noindent Volker, S., \& Hernquist, L. 2005, ApJ, 622, L9

\noindent Walker, I.R., Mihos, J.C., \& Hernquist, L. 1996, ApJ,
460, 121

\noindent Wielen, R. 1990, Dynamics and interactions of galaxies
(Berlin: Springer-Verlag)

\noindent Wright, G.S., James, P.A., Joseph, R.D., \& McLean, I.S.
1990, Nature, 344, 417

\noindent Wozniak, H., Friedli, D., Martinet, L., Martin, P., \&
Bratschi, P.   1995, A \& A S, 111, 115

\end{document}